\documentclass[usegraphicx,usenatbib]{mn2e}
\usepackage{amsmath}
\usepackage{amssymb}
\usepackage{times}
\usepackage{subfigure}
\usepackage{epsfig}

\renewcommand{\descriptionlabel}[1]%
 {\hspace{\labelsep}\textbf{#1}}

\title[Upper Limits on the Hot Jupiter Fraction in the Field of NGC~7789]
  {Upper Limits on the Hot Jupiter Fraction in the Field of NGC~7789}

\author[D.M. Bramich and Keith Horne]
  {D.M.~Bramich$^{1,2}$\thanks{E-mail: dmb7@st-and.ac.uk}
   and Keith~Horne$^1$
 \medskip
 \\$^1$School of Physics \& Astronomy,
   University of St.~Andrews, North Haugh,
   St.~Andrews, Fife,
   KY16 9SS, UK
 \\$^2$Astrophysics Research Institute, 
   Liverpool John Moores University,
   Twelve Quays House,
   Egerton Wharf,
   Birkenhead,
   CH41 1LD, UK}

\begin{document}

\date{Accepted 2005 August ???. Received 2005 August ???; Submitted 2005 Nov 4}

\pagerange{\pageref{firstpage}--\pageref{lastpage}} \pubyear{2004}

\maketitle

\label{firstpage}

\begin{abstract} 
We describe a method of estimating the abundance of short-period extrasolar planets based on the results of a 
photometric survey for planetary transits. We apply the method to a 21-night survey with the 2.5m Isaac Newton
Telescope of $\sim$32000 stars in a $\sim 0.5\degr \times 0.5\degr$ square field including the open cluster NGC~7789. From the 
colour-magnitude diagram we estimate the mass and
radius of each star by comparison with the cluster main sequence. 
We search for injected synthetic transits throughout the lightcurve of each star in order to determine their recovery rate,
and thus calculate the expected number of transit detections and false alarms in the survey.
We take proper account of the photometric accuracy, time sampling of the observations and criteria
(signal-to-noise and number of transits) adopted for transit detection. Assuming that none of the transit
candidates found in the survey will be confirmed as real planets, we place conservative upper limits
on the abundance of planets as a function of planet radius, orbital period and spectral type.
\end{abstract} 


\section{Introduction}

Photometric surveys for transiting extra-solar planets have become very popular since the detection of 
the transits exhibited by the planet-host star HD 209458 (\citealt{cha00}; \citealt{bro01}). For the first 
time the radius of an extra-solar planet was determined, and the measurement of the orbital inclination 
lead to an estimate of the planetary mass, not just a lower limit. The planet HD 209458b was found to 
have an average density of $\sim$0.38 g/cm$^3$, significantly less than the average density of Saturn 
(0.7 g/cm$^3$), leading to the term ``hot Jupiter'' for the class of Jupiter mass planets with 
short periods (1-10 d). Transiting planets are also very important in that their atmospheric 
composition may be determined from transmission spectroscopy for the brighter host stars 
(\citealt{cha02}; \citealt{bro02}; \citealt{vid03}). Careful modelling of the transit morphology and/or  
timings may be used to constrain the presence of moons or rings and to probe the limb darkening 
of the star (\citealt{bro01}). 

Since the discovery of the transiting nature of HD 209458b, many transit candidates have been put 
forwards by various groups (e.g: \citealt{str03}; \citealt{dra04}; \citealt{bra05}). 
OGLE have been by far the most prolific transit survey with 177 transit candidates from three observational seasons 
(\citealt{uda02a}; \citealt{uda02b}; \citealt{uda03}; \citealt{uda04}). However, even with the 
discovery of numerous candidates, follow-up observations have confirmed the planetary status of only six, 
bringing the total number of transiting planets to nine (see \citealt{bra05} and references therein; \citealt{sat05};
\citealt{bou05}). 
This is due to the ubiquity of eclipsing binaries and the many observational scenarios 
involving these systems that mimic a transit event (\citealt{bro03}). Spectroscopic and multi-band photometric observations are 
required to rule out the eclipsing binary scenarios and determine the mass of the companion (e.g: \citealt{alo04}).

When hunting for new planets, 
the main advantage of the transit method over the radial-velocity technique is that many stars may be 
monitored in parallel and to fainter magnitudes, thus probing out beyond the Solar neighbourhood.
Even though only a small fraction ($\sim$0.1\%) of stars are expected to exhibit a hot Jupiter transit signal, 
by using a large field of view instrument on a crowded star field one can monitor enough stars
to the precision required to detect a number of transiting planets. Consequently large charge-coupled device (CCD)
mosaic cameras are essential to the planet catch potential of a transit survey.

A transit survey produces transit candidates that need follow-up observations to determine 
the nature of the transit signals. Candidates confirmed as transiting planets add to our database of extra-solar planets and 
constrain their poorly known mass-radius relationship (\citealt{bur04}). To estimate the fraction of stars that harbour a planet
(the planet fraction) as a function of spectral type and planet type we compare the number of transiting planets
detected with a calculation of the expected number of transiting planet detections.
Even when zero planets are detected (a null result) such a calculation places upper limits on the planet fraction.

\begin{table*}
\centering
\caption{The different subsets of stars used when calculating the expected number of
         transiting planet detections and false alarms. \label{tab:spectypes}}
\begin{tabular}{lcrr}
\hline
Set Of Stars & Mass Range & No. Of Stars & No. Of Stars With 1999-07 Lightcurve Data \\
\hline
All Stars    & $0.08 M_{\sun} \leq M_{*} \leq 1.40 M_{\sun}$ & 32027 & 20949 \\
Late F Stars & $1.05 M_{\sun} \leq M_{*} \leq 1.40 M_{\sun}$ & 3129  & 2780  \\
G Stars      & $0.80 M_{\sun} \leq M_{*} \leq 1.05 M_{\sun}$ & 7423  & 6711  \\
K Stars      & $0.50 M_{\sun} \leq M_{*} \leq 0.80 M_{\sun}$ & 15381 & 9690  \\
M Stars      & $0.08 M_{\sun} \leq M_{*} \leq 0.50 M_{\sun}$ & 6094  & 1768  \\
\hline
\end{tabular}
\end{table*}

In this paper we describe a Monte Carlo method for calculating detection probabilities 
(and false alarm rates) of transiting planets based on photometric data, as 
functions of various parameters, taking into account the following factors:
\begin{enumerate}
  \item Limb darkening effects which tend to make central eclipses deeper and grazing eclipses shallower.
  \item The effect of orbital inclination on the shape and width of the transit lightcurve.
  \item The distribution of the photometric data in time and the individual error bars on each measurement.
  \item The signal-to-noise threshold, number of transits, and number of data points in-transit and out-of-transit required
        for a detection. 
\end{enumerate}
We then apply the method to the transit survey described in \citealt{bra05} (here on referred to as BRA05) to determine
the expected number of transiting planet detections and place limits on the planet fraction as a function of 
star and planet type.

In Section 2 we describe the lightcurve data used in the analysis and in Section 3 we define the detection probabilities and false alarm 
probabilities for an extra-solar planet based on photometric data. In Section 4 we present the Monte Carlo method 
that we used to calculate these probabilities and derive limits on the hot Jupiter fraction in the field of NGC~7789 as a 
function of star and planet type. In Section 5 we discuss the results and in Section 6 we present our conclusions.

\section{The Transit Survey Data On NGC~7789}

\begin{figure}
\epsfig{file=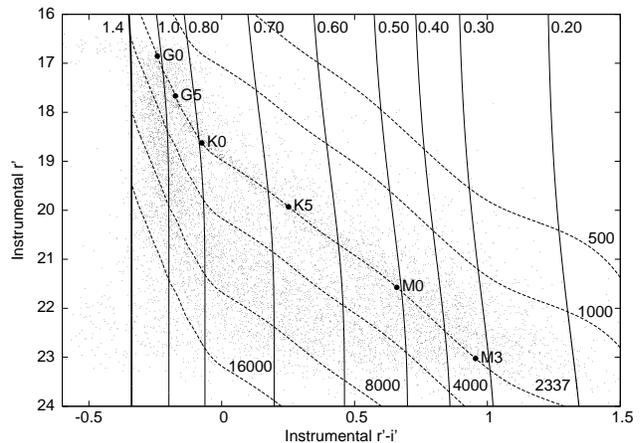,angle=270.0,width=\linewidth}
\caption{Instrumental CMD for chip 4 (taken from BRA05) which was centred on the cluster. The numbers
         along the top are masses in units of $M_{\sun}$ and the numbers along the bottom and right are 
         distances in parsecs. Fiducial spectral types are marked on the cluster theoretical main sequence
         for clarity.
\label{fig:CMD}}
\end{figure}

A transit survey of the field of NGC~7789 was presented in BRA05 in which 
$\sim$33000 stars were photometrically monitored in the Sloan $r^{\prime}$ band over three separate runs 
with dates 1999 June 22-30,
1999 July 22-31 and 2000 September 10-20. For brevity these runs shall be refered to from now on as 
1999-06, 1999-07 and 2000-09 respectively. 

To summarise, in BRA05, Sloan $r^{\prime}-i^{\prime}$ colour 
indices were used to construct a colour-magnitude diagram (CMD) and thereby identify the cluster 
main sequence. Fig.~\ref{fig:CMD} shows the CMD for
the stars from chip 4 which was centred on the cluster. Although the cluster main sequence is visible, it is clear
that most stars in the sample are field stars and not cluster stars.
A theoretical main sequence model for the stellar mass range 
$0.08~M_{\sun}~\leq~M_{*}~\leq~1.40~M_{\sun}$ was adopted and fitted to the cluster main sequence via 
magnitude offsets. Using the known cluster distance $d_{\mbox{\small c}} = 2337$pc and reddening 
$E(B~-~V)~=~0.217$, and adopting an Einasto law for the 
distribution of the interstellar medium in the Milky Way (\citealt{rob03}), a distance $d_{*}$ was 
determined for each star such that the 
theoretical main sequence passes through the star's position on the CMD.
It was argued that giant stars lie beyond the edge of the 
galaxy in order to be non-saturated in the image data. Hence, it was assumed that each star is on the main sequence,
and after determining the star's distance $d_{*}$, the star's mass $M_{*}$ and radius $R_{*}$ 
could be read off from its position on the theoretical main sequence. 

In this paper we consider the 32027 stars from this data set that have a lightcurve from the 2000-09 run, 
and an assigned distance, mass and radius. The remaining stars with lightcurves lack a colour 
measurement or were too blue to be assigned a mass and radius using the adopted theoretical main sequence. We 
also consider the lightcurve data from the 1999-07 run 
where it exists. BRA05 searched for transits in the 10-night 1999-07 run and the 11-night 2000-09 run, 14 months later. 
The 1999-06 run was too sparsely sampled in time to support transit hunting by the adopted search technique. 

We are interested in the expected number of transiting planet detections (and false alarms) for 
stars of different masses or, equivalently, spectral types. To facilitate this analysis we consider 4 
mutually exclusive subsets of stars which in union make up the set of all 32027 stars. These sets are the late 
F stars, G stars, K stars and M stars respectively. Table \ref{tab:spectypes} 
shows the number of stars in each set and 
the spectral type/mass ranges to which they correspond. The table also includes the number of 
stars for which 1999-07 lightcurve data exists. The mass ranges for the various spectral 
types 
are taken from \citet{lan92}. 

\section{Detection Probabilities and False Alarm Rates}

In BRA05, a matched filter algorithm was used to search for transits in the lightcurves by adopting a 
square ``boxcar''
shape for the transit lightcurve of total width $5 \Delta t$ (where $\Delta t$ is the transit duration
searched for). This search was based on the transit detection statistic:
\begin{equation}
S_{\mbox{\small tra}}^{2} \equiv \frac{ \chi^{2}_{\mbox{\small const}} - \chi^{2}_{\mbox{\small tra}} }
            { \left(\frac{ \displaystyle \chi^{2}_{\mbox{\small out}} }
                { \displaystyle N_{\mbox{\small out}} - 1 }\right) } \label{eqn:trastat}
\end{equation}
where $\chi^{2}_{\mbox{\small tra}}$ is the chi squared of the boxcar transit fit, $\chi^{2}_{\mbox{\small const}}$
is the chi squared of the constant fit, $\chi^{2}_{\mbox{\small out}}$ is the chi squared of
the boxcar transit fit for the $N_{\mbox{\small out}}$ out-of-transit data points.
Transit candidates were chosen using a threshold of $S_{\mbox{\small tra}}~\geq~S_{\mbox{\small min}}~=~10$.

Consider an extra-solar planet of radius $R_{\mbox{\small p}}$, orbital period $P$ and orbital inclination $i$
with $t_{0}$ as the time of mid-transit. The planet orbits a star S, of known mass $M_{*}$ and
radius $R_{*}$, that has an associated lightcurve. We calculate the predicted transit lightcurves
based on a simple planet-star model: we assume a luminous primary, linear limb darkening with $u = 0.5$ and a dark massless 
companion in a circular orbit. Adding this signal into the 
observed lightcurve of the star, we calculate the transit statistic $S_{\mbox{\small tra}}$ (Eqn.~\ref{eqn:trastat}) 
for each transit event, and then evaluate the following detection function:
\begin{multline}
D \left(\text{S},S_{\mbox{\small min}},N_{\mbox{\small min}},R_{\mbox{\small p}},P,i,t_{0}\right) \\ =
  \begin{cases}
    1 & \text{if $S_{\mbox{\small tra}} \geq S_{\mbox{\small min}}$ for at least
              $N_{\mbox{\small min}}$ predicted transits} \\
    0 & \text{otherwise}
  \end{cases} 
\label{eqn:detfunc} 
\end{multline}
Using the same procedure as above, but without actually adding the predicted transit lightcurve into the
observed lightcurve of the star, we evaluate the false alarm function:
\begin{multline}
F \left(\text{S},S_{\mbox{\small min}},N_{\mbox{\small min}},R_{\mbox{\small p}},P,i,t_{0}\right) \\ =
  \begin{cases}
    1 & \text{if $S_{\mbox{\small tra}} \geq S_{\mbox{\small min}}$ for at least
              $N_{\mbox{\small min}}$ predicted transits} \\
    0 & \text{otherwise}
  \end{cases}
\label{eqn:falfunc}
\end{multline}

The function $D$ is a trigger function that indicates where the data and detection algorithm are capable
of detecting a transit of the specified type, and $F$ indicates where the data alone suggest that such a transit is
actually present. The BRA05 lightcurve data contains some eclipsing binary stars and possibly transits. Hence both $D$ and $F$
are slightly over estimated.

\begin{figure}
\epsfig{file=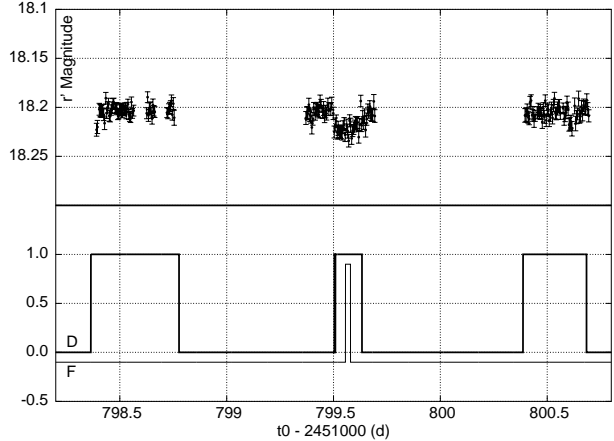,angle=270.0,width=\linewidth}
\caption{{\bf Upper panel:} A subsection of the lightcurve of star 61377 with an injected 0.02~mag offset of duration
         3~h starting at $t_{0} = 2451799.5$.
         {\bf Lower panel:} The periodic functions $D \left(t_{0}\right)$ and
         $F \left(t_{0}\right)$ for star 61377 represented by thick and thin
         continuous lines respectively. $F \left(t_{0}\right)$ is offset vertically from $D \left(t_{0}\right)$ by 0.1 for clarity.
         We adopted $S_{\mbox{\scriptsize min}}~=~10$, $N_{\mbox{\scriptsize min}}~=~1$,
         $R_{\mbox{\scriptsize p}}~=~1.2 R_{\mbox{\scriptsize J}}$, $P~=~3.338$~d and $i~=~90\degr$ for this calculation.
\label{fig:DSFS}}
\end{figure}

In the upper panel of Fig.~\ref{fig:DSFS} we plot a subsection of the lightcurve of star 61377 with an injected 0.02~mag offset of duration
3~h starting at $t_{0} = 2451799.5$~d. This
$r^{\prime} \approx 18.20$ mag G star has a mass, radius and distance of 0.96$M_{\sun}$, 0.96$R_{\sun}$ and 3152pc respectively.
In the lower panel of Fig.~\ref{fig:DSFS} we plot the corresponding periodic functions $D \left(t_{0}\right)$ and
$F \left(t_{0}\right)$ represented by thick and thin
continuous lines respectively. We adopted $S_{\mbox{\small min}}~=~10$, $N_{\mbox{\small min}}~=~1$,
$R_{\mbox{\small p}}~=~1.2 R_{\mbox{\small J}}$, $P~=~3.338$~d and $i~=~90\degr$ for this calculation. The function $D \left(t_{0}\right)$
attains the value of 1 where there is data of sufficient precision to detect a transit.
The function $F \left(t_{0}\right)$ attains the value of 1 where there is data that mimics a transit
signature, and it has clearly been triggered by the injected offset in the lightcurve data.

In Eqn.~\ref{eqn:detfunc}, we integrate out the ``nuisance parameters'' $t_{0}$ and $i$ to get:
\begin{multline}
\text{P}
\left(\text{det}\,|\,\text{S},S_{\mbox{\small min}},N_{\mbox{\small min}},R_{\mbox{\small p}},P\right) \\ =
\int_{0\degr}^{90\degr} \text{d}i \int_{0}^{P} \text{d}t_{0} f(t_{0},i)
D \left(\text{S},S_{\mbox{\small min}},N_{\mbox{\small min}},R_{\mbox{\small p}},P,i,t_{0}\right)
\label{eqn:pdet1}
\end{multline}
where
P$\left(\text{det}\,|\,\text{S},S_{\mbox{\small min}},N_{\mbox{\small min}},R_{\mbox{\small p}},P\right)$
is the detection probability for star S and $f(t_{0},i)$ is the joint probability distribution function (PDF) of $t_0$ and $i$.
We assume that the parameters $t_{0}$ and $i$ are independent, $t_{0}$ is uniformly distributed
over $0~\leq~t_{0}~\leq~P$ and random orbit orientation for $i$ in the range $0\degr \leq i \leq 90\degr$.
Hence we may write:
\begin{equation}
f(t_{0},i) = \frac{\pi \sin i}{180\degr P} \label{eqn:indep}
\end{equation}
Combining Eqns. \ref{eqn:pdet1} and \ref{eqn:indep} we get the detection probability:
\begin{multline}
\text{P}
\left(\text{det}\,|\,\text{S},S_{\mbox{\small min}},N_{\mbox{\small min}},R_{\mbox{\small p}},P\right) \\ =
\int_{0\degr}^{90\degr} \text{d}i \int_{0}^{P} \text{d}t_{0} \left( \frac{\pi \sin i}{180\degr P} \right)
D \left(\text{S},S_{\mbox{\small min}},N_{\mbox{\small min}},R_{\mbox{\small p}},P,i,t_{0}\right)
\label{eqn:pdet}
\end{multline}
Using a parallel argument, we obtain an expression for the false alarm probability
P$\left(\text{fal}\,|\,\text{S},S_{\mbox{\small min}},N_{\mbox{\small min}},R_{\mbox{\small p}},P\right)$
as:
\begin{multline}
\text{P}
\left(\text{fal}\,|\,\text{S},S_{\mbox{\small min}},N_{\mbox{\small min}},R_{\mbox{\small p}},P\right) \\ =
\int_{0\degr}^{90\degr} \,\text{d}i \int_{0}^{P} \,\text{d}t_{0}
\left( \frac{\pi \sin i}{180\degr P} \right)
F \left(\text{S},S_{\mbox{\small min}},N_{\mbox{\small min}},R_{\mbox{\small p}},P,i,t_{0}\right)
\label{eqn:pfal}
\end{multline}

\section{Monte Carlo Simulations}

\subsection{Methodology}

\begin{figure*}
\def\subfigtopskip{4pt}
\def\subfigbottomskip{8pt}
\def\subfigcapskip{4pt}
\centering
\begin{tabular}{cc}
\subfigure[]{\epsfig{file=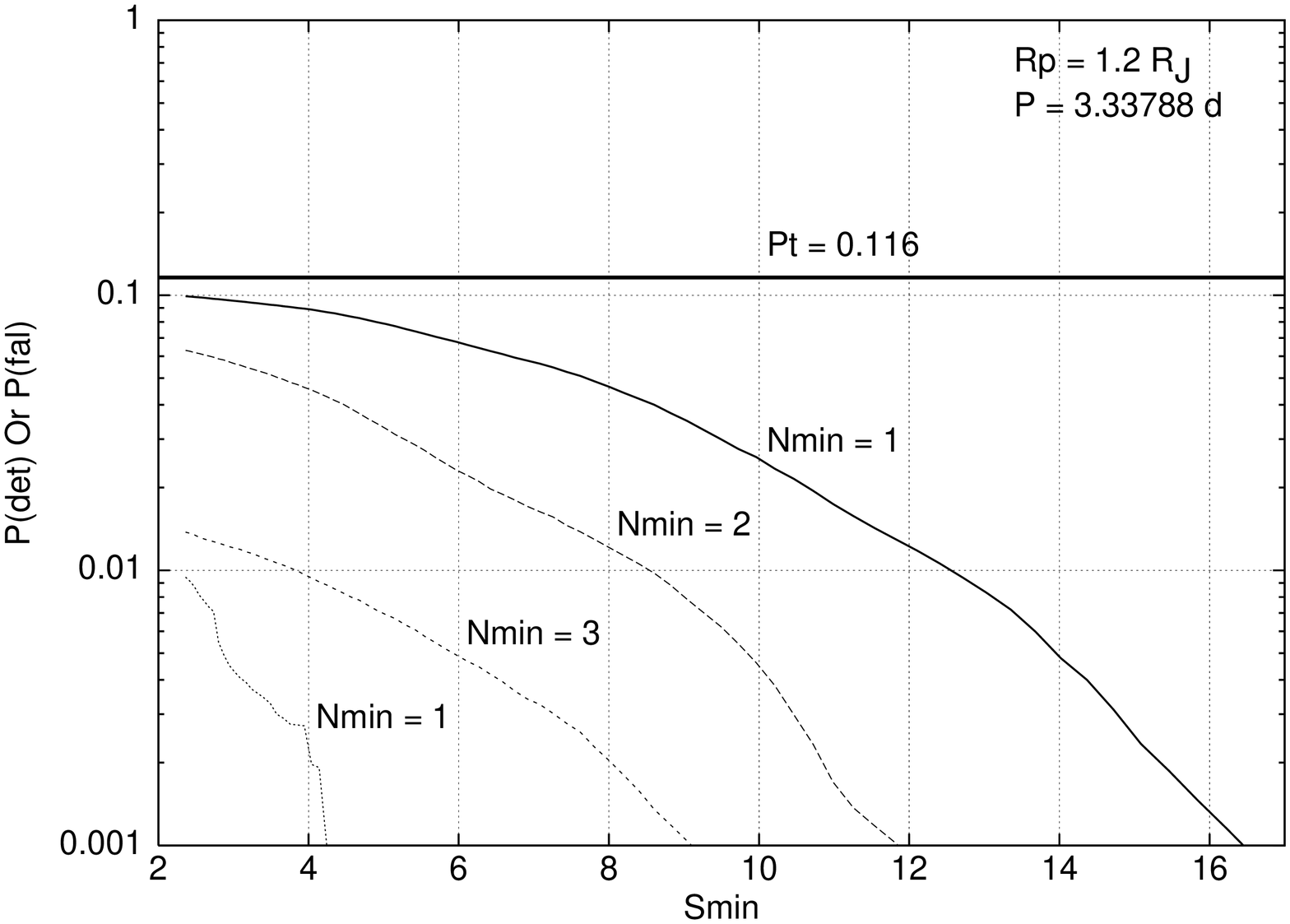,angle=270.0,width=0.5\linewidth} \label{fig:lc61377a}} &
\subfigure[]{\epsfig{file=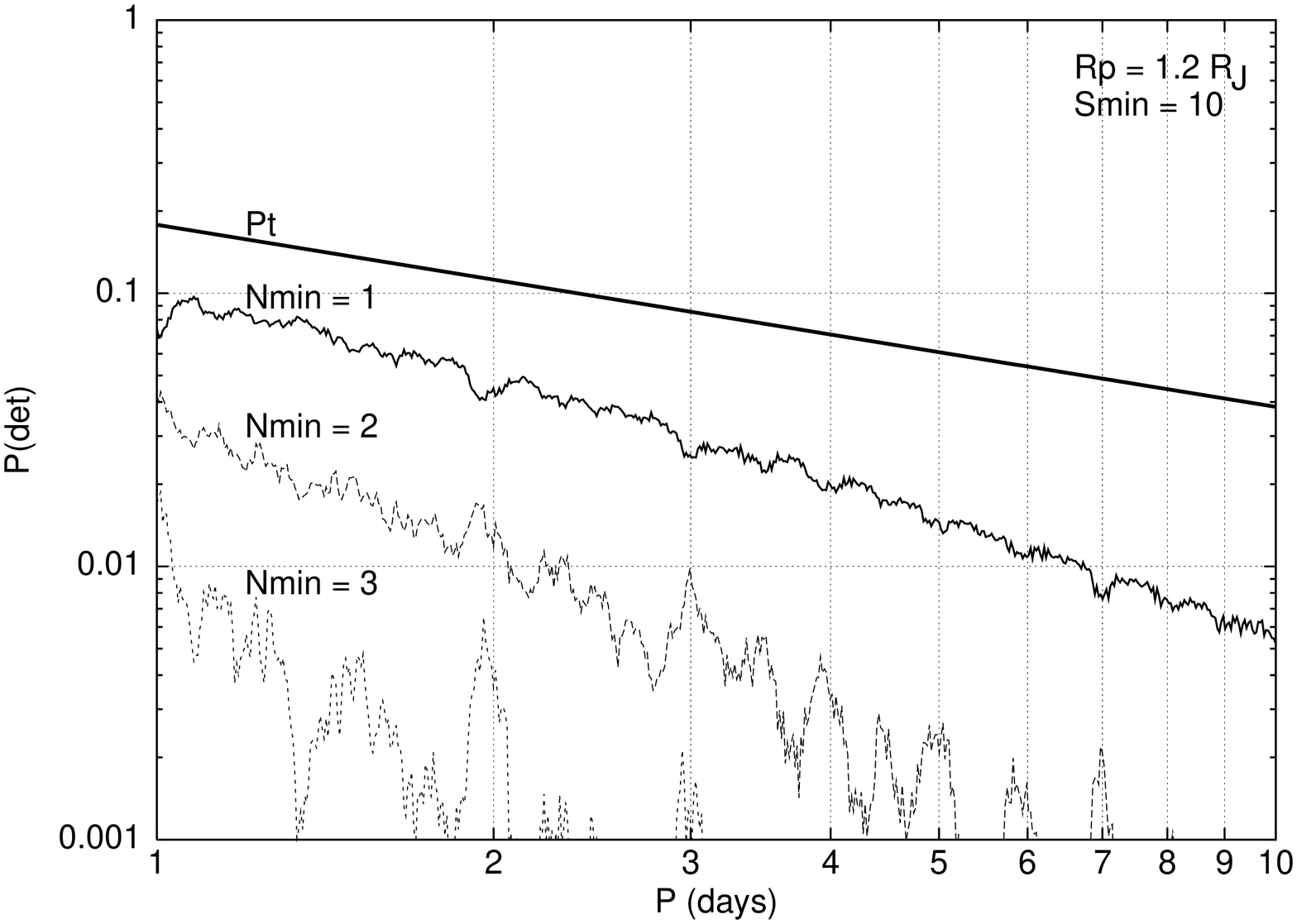,angle=270.0,width=0.5\linewidth} \label{fig:lc61377b}} \\
\end{tabular}
\caption{(a): Detection probability 
         as a function of $S_{\mbox{\scriptsize min}}$ for star 61377 with $P~=~3.338$~d and 
         $R_{\mbox{\scriptsize p}} = 1.2 R_{\mbox{\scriptsize J}}$. The continuous, dashed and 
         shorter dashed lines correspond to $N_{\mbox{\scriptsize min}}~=~1$, $N_{\mbox{\scriptsize min}}~=~2$
         and $N_{\mbox{\scriptsize min}}~=~3$ respectively. The false alarm probability 
         as a function of $S_{\mbox{\scriptsize min}}$ for the same star and planet with $N_{\mbox{\scriptsize min}}~=~1$
         is shown by the dotted line. The thick continuous line corresponds to the probability 
         $\text{P}_{\mbox{\scriptsize t}}~=~0.116$ that the planet-star system exhibits transits. 
         (b): Detection probability as a function of
         $P$ for star 61377 with $S_{\mbox{\scriptsize min}}~=~10$ and $R_{\mbox{\scriptsize p}} = 1.2 R_{\mbox{\scriptsize J}}$.
         Again, the continuous, dashed and shorter dashed lines correspond to $N_{\mbox{\scriptsize min}}~=~1$, 
         $N_{\mbox{\scriptsize min}}~=~2$ and $N_{\mbox{\scriptsize min}}~=~3$ respectively.
         The false alarm probability is approximately zero for all $P$ and $N_{\mbox{\scriptsize min}}$
         at this detection threshold. The thick continuous curve corresponds to the probability $\text{P}_{\mbox{\scriptsize t}}$
         that the planet-star system exhibits transits.
\label{fig:lc61377}}
\end{figure*}

We take the Monte Carlo approach to evaluating the detection probabilities and false alarm rates (\citealt{pre92}),
rather than attempting to numerically integrate Eqns.~\ref{eqn:pdet} and~\ref{eqn:pfal}. In general, a Monte Carlo 
simulation estimates the probability of an event by selecting a large random sample from the
parameter space as governed by the underlying PDF, and then calculating the fraction of the sample that satisfy
the event criteria. The larger the sample, the more accurate the calculated probability. However, the size of the
sample that may be selected and analysed is usually limited by available computing resources.

For each star S and its corresponding lightcurve, we used the Monte Carlo method to calculate
P$\left(\text{det}\,|\,\text{S},S_{\mbox{\small min}},N_{\mbox{\small min}},R_{\mbox{\small p}},P\right)$
and
P$\left(\text{fal}\,|\,\text{S},S_{\mbox{\small min}},N_{\mbox{\small min}},R_{\mbox{\small p}},P\right)$
for a grid in $S_{\mbox{\small min}}$, $N_{\mbox{\small min}}$, $R_{\mbox{\small p}}$ and $P$.
We chose to use a geometric sequence in $S_{\mbox{\small min}}$ from $S_{\mbox{\small min}}~=~2.4$ to
$S_{\mbox{\small min}}~=~26.4$ with geometric factor 1.025. We also chose a geometric sequence in $P$ from $P = 1$~d
to $P = 10$~d with geometric factor 1.004 resulting in 576 period values.
We chose $N_{\mbox{\small min}}~\in~\{ 1,2,3 \}$ and
$R_{\mbox{\small p}}~=~1.2~R_{\mbox{\small J}}$. For each grid point we selected a set of
$N_{\mbox{\footnotesize MC}} = 1000$ planets with $t_{0}$ and $i$ drawn randomly for each planet from the
PDF in Eqn.~\ref{eqn:indep}.

The grid for $P$ should be fine enough that the difference in period $\Delta~P~=~P_{i+1}~-~P_{i}$ between two
consecutive grid points $P_{i}$ and $P_{i+1}$ (where $P_{i+1} > P_{i}$) is such that the difference in the number
of cycles spanning the duration of the lightcurve is less than or equal to a fraction $f_{\mbox{\small t}}$ of the
transit duration (in cycle units). This condition implies that the grid in $P$ should be a geometric sequence
with geometric factor less than or equal to $1 + (f_{\mbox{\small t}} \Delta t / T)$ where
$\Delta t$ is the transit duration and $T$ is the duration of the lightcurve.
Adopting $f_{\mbox{\small t}} = 0.5$, $\Delta t \approx 2$~h for a typical transit
duration and $T = 10.4$~d corresponding to the longer 2000-09 run yields
$f_{\mbox{\small t}} \Delta t / T \approx 4.0 \times 10^{-3}$. Hence our choice of grid in $P$ is fine enough
for the $\sim$35\% of stars that have lightcurve data from the 2000-09 run alone. For the remaining stars with
lightcurve data from both runs, adopting the much finer period grid that is required makes the Monte Carlo simulations
prohibitive in terms of computer processing time. To fully sample the possible period aliases introduced by using the adopted
grid, the current period value $P_{\mbox{\small curr}}$ for each Monte Carlo trial was drawn from a uniform distribution on
the interval $P/\sqrt{1.004} \le P_{\mbox{\small curr}} \le P\sqrt{1.004}$.

For the 1-10 d planets that we consider, the probability that a planet transits is $\sim$10\%. Hence, the number of Monte Carlo
trials that result in a detection is smaller than $N_{\mbox{\small MC}}$ by at least a factor of 10, leading to a relatively
noisy determination of the detection probabilities (for $N_{\mbox{\small MC}} = 1000$, $\sigma \sim$10\%). However, when
we sum the detection probabilities over a large number of stars, typically $\sim~10^{4}$ (see Section 4.3), the noise is
reduced to an insignificant level ($\sim$0.1\%). Similarly, it is reduced even further when we integrate over various period ranges 
(see Section 4.4).

\subsection{An Example Simulation}

\begin{figure*}
\def\subfigtopskip{4pt}   
\def\subfigbottomskip{8pt}
\def\subfigcapskip{4pt}
\centering
\begin{tabular}{cc}
\subfigure[]{\epsfig{file=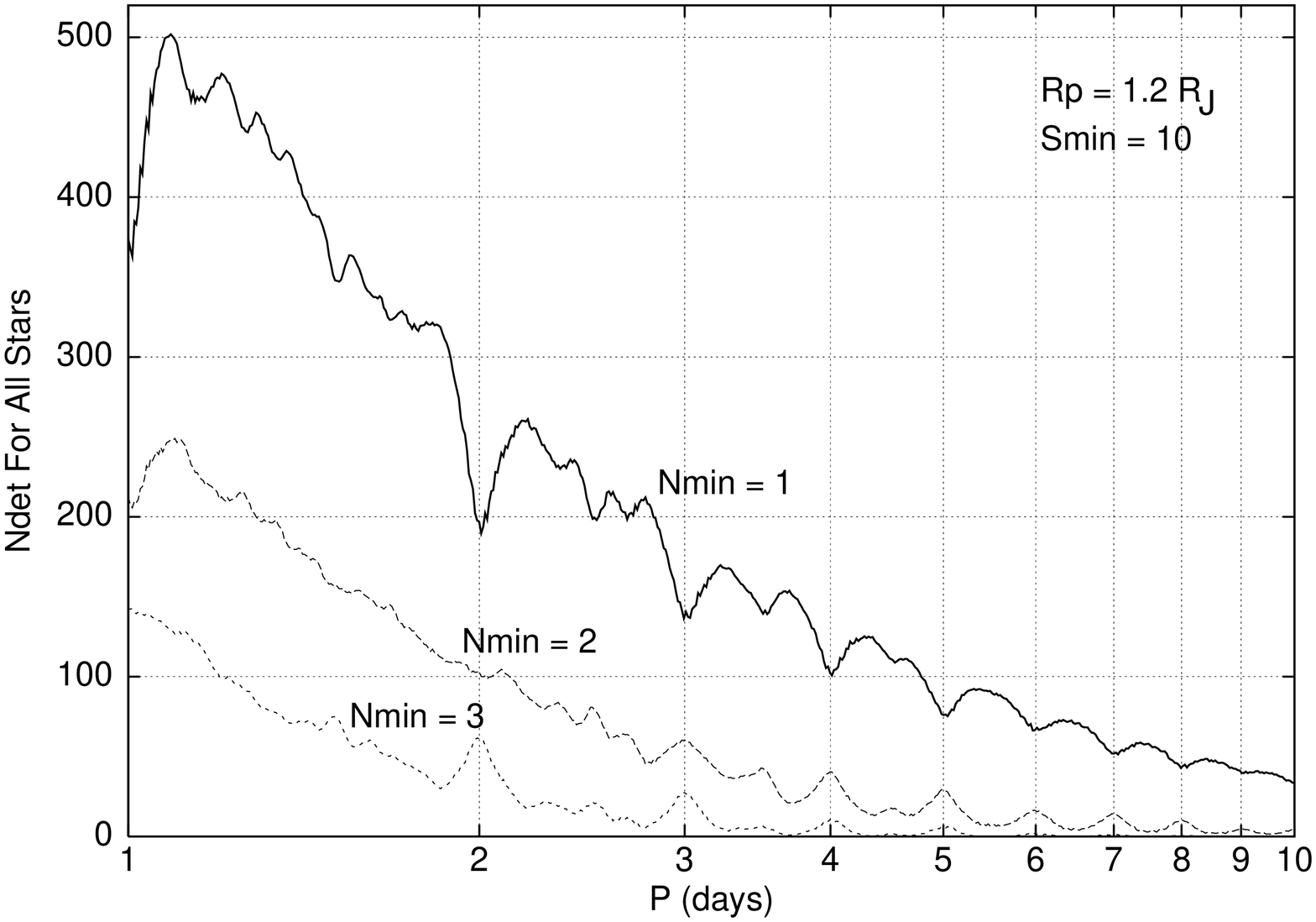,angle=270.0,width=0.5\linewidth} \label{fig:Ndet_vs_P}} & 
\subfigure[]{\epsfig{file=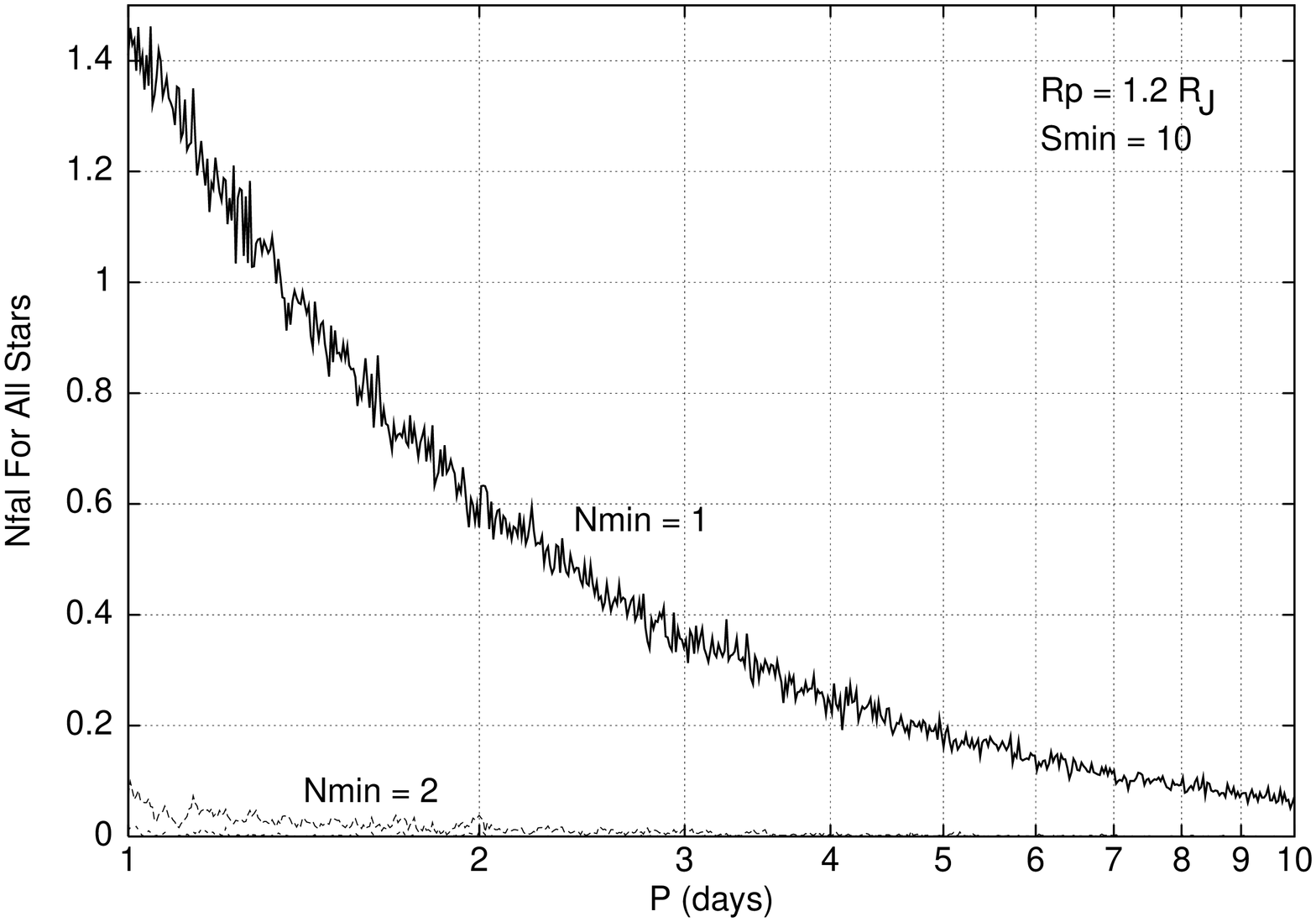,angle=270.0,width=0.5\linewidth} \label{fig:Nfal_vs_P}} \\
\end{tabular}
\caption{(a): Expected number of transiting planet detections for all stars as a function of $P$ for
         $S_{\mbox{\scriptsize min}}~=~10$ and $R_{\mbox{\scriptsize p}} = 1.2 R_{\mbox{\scriptsize J}}$.
         The continuous, dashed and
         shorter dashed lines correspond to $N_{\mbox{\scriptsize min}}~=~1$, $N_{\mbox{\scriptsize min}}~=~2$
         and $N_{\mbox{\scriptsize min}}~=~3$ respectively.
         (b): Expected number of false alarms for all stars as a function of $P$ for
         $S_{\mbox{\scriptsize min}}~=~10$ and $R_{\mbox{\scriptsize p}} = 1.2 R_{\mbox{\scriptsize J}}$.
         Again, the continuous, dashed and
         shorter dashed lines correspond to $N_{\mbox{\scriptsize min}}~=~1$, $N_{\mbox{\scriptsize min}}~=~2$
         and $N_{\mbox{\scriptsize min}}~=~3$ respectively.
\label{fig:NdetNfal_vs_P}}
\end{figure*}

Let us now consider the complete lightcurve of star 61377.
The star has 118 data points over 10 nights in its 1999-07 lightcurve with a standard deviation of $\sim$0.007~mag and 612 data
points over 11 nights in its 2000-09 lightcurve with a standard deviation of $\sim$0.010~mag.

In Fig. \ref{fig:lc61377}, we plot the detection probability and false alarm probability as functions of the
transit statistic detection threshold (Fig.~\ref{fig:lc61377a}) and period (Fig.~\ref{fig:lc61377b}) for star 61377. 
We used $N_{\mbox{\small MC}}~=~10^{5}$ Monte-Carlo trials for this star in order to reduce the noise from the 
simulations for illustrative purposes. 
The detection and false alarm probabilities both decrease with detection threshold $S_{\mbox{\small min}}$
(Fig. \ref{fig:lc61377a}). For this particular star, it can be seen that false alarms are very unlikely even at very low detection
thresholds. The thick continuous line corresponds to the probability $\text{P}_{\mbox{\small t}}~=~0.116$ that the planet-star 
system exhibits transits, calculated from: 
\begin{align}
\text{P}_{\mbox{\small t}} &= \frac{R_{\mbox{\small p}} + R_{*}}{a} \notag \\
                           &= 0.162 \left(\frac{R_{\mbox{\small p}} + R_{*}}{R_{\sun}}\right) 
                                    \left(\frac{M_{*}}{M_{\sun}}\right)^{-1/3} 
                                    \left(\frac{P}{1 \text{d}}\right)^{-2/3} \notag \\  
\label{eqn:Ptra}
\end{align}
where $a$ is the orbital radius and the final expression uses Kepler's law. 
The completeness of the transit search falls rapidly with the adopted detection threshold
$S_{\mbox{\small min}}$. For $N_{\mbox{\small min}} = 1$, the completeness is $\sim$76.7\% at threshold 
$S_{\mbox{\small min}}~=~4$, dropping to $\sim$22.3\% for $S_{\mbox{\small min}}~=~10$.

In Fig. \ref{fig:lc61377b} we see the expected $P^{-2/3}$ dependence of detection probability on orbital period, but with 
more detailed period structure arising from the detailed time sampling of the observations.
For $N_{\mbox{\small min}} = 1$, orbital periods close to integer values tend to have lower detection probabilities since such periods
are resonant with the observational gaps during the daytime. Conversely, orbital periods close to fractional values tend to 
have higher detection probabilities since such periods cover a greater range of orbital phases. For example, periods close to 
$\sim$3.0~d have a detection probability of $\sim$0.025 whereas periods close to $\sim$2.7~d have a detection probability of
$\sim$0.036 for this particular star. Fig. \ref{fig:lc61377b} also shows that as you increase the number of recovered transits required 
for a detection, the detection probability decreases rapidly.

\subsection{Expected Number Of Transiting Planet Detections}

\begin{figure*}
\def\subfigtopskip{4pt}
\def\subfigbottomskip{8pt}
\def\subfigcapskip{4pt}
\centering
\begin{tabular}{cc}
\subfigure[]{\epsfig{file=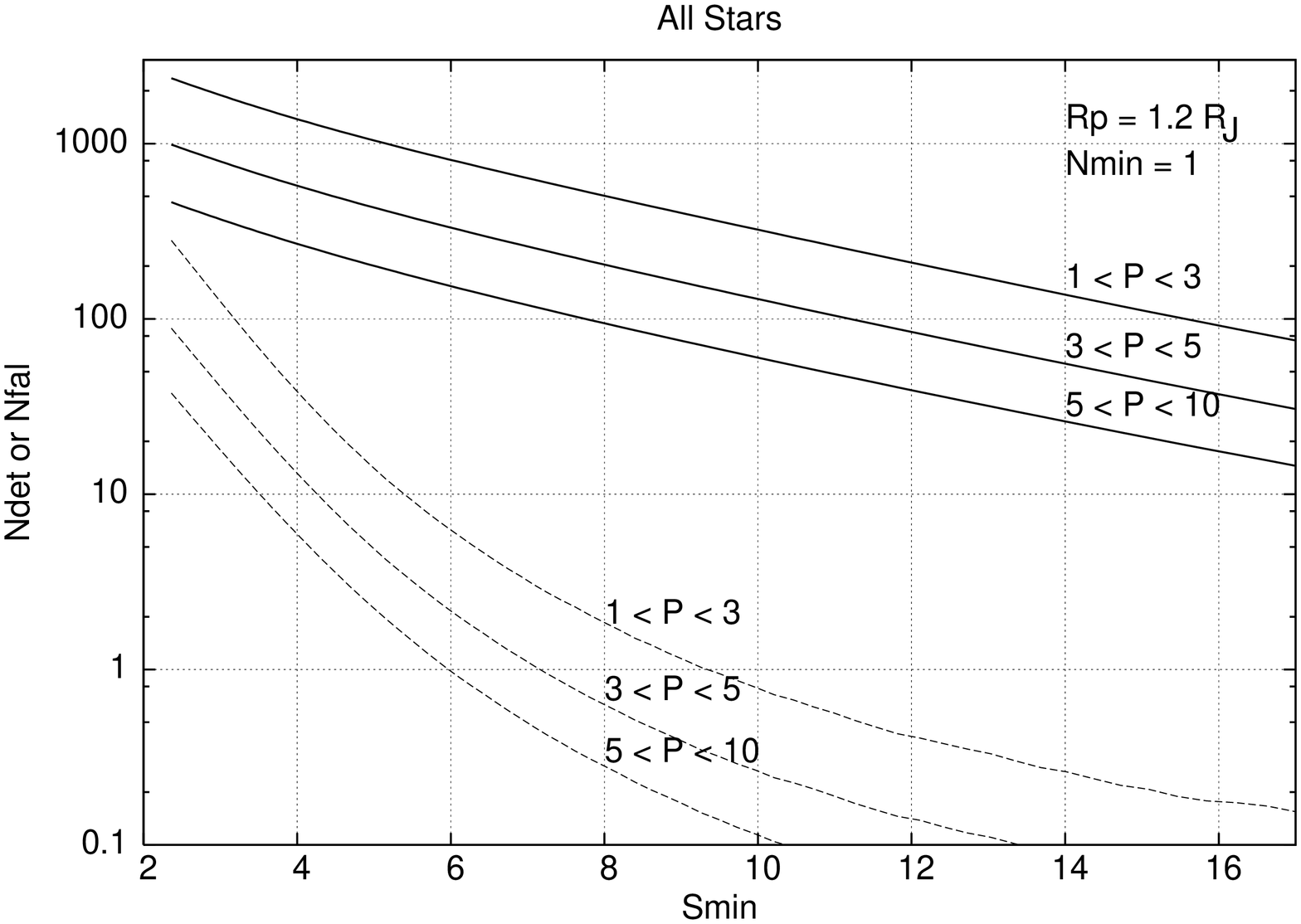,angle=270.0,width=0.5\linewidth} \label{fig:Ndet_vs_thresh_Allstar}} &
\subfigure[]{\epsfig{file=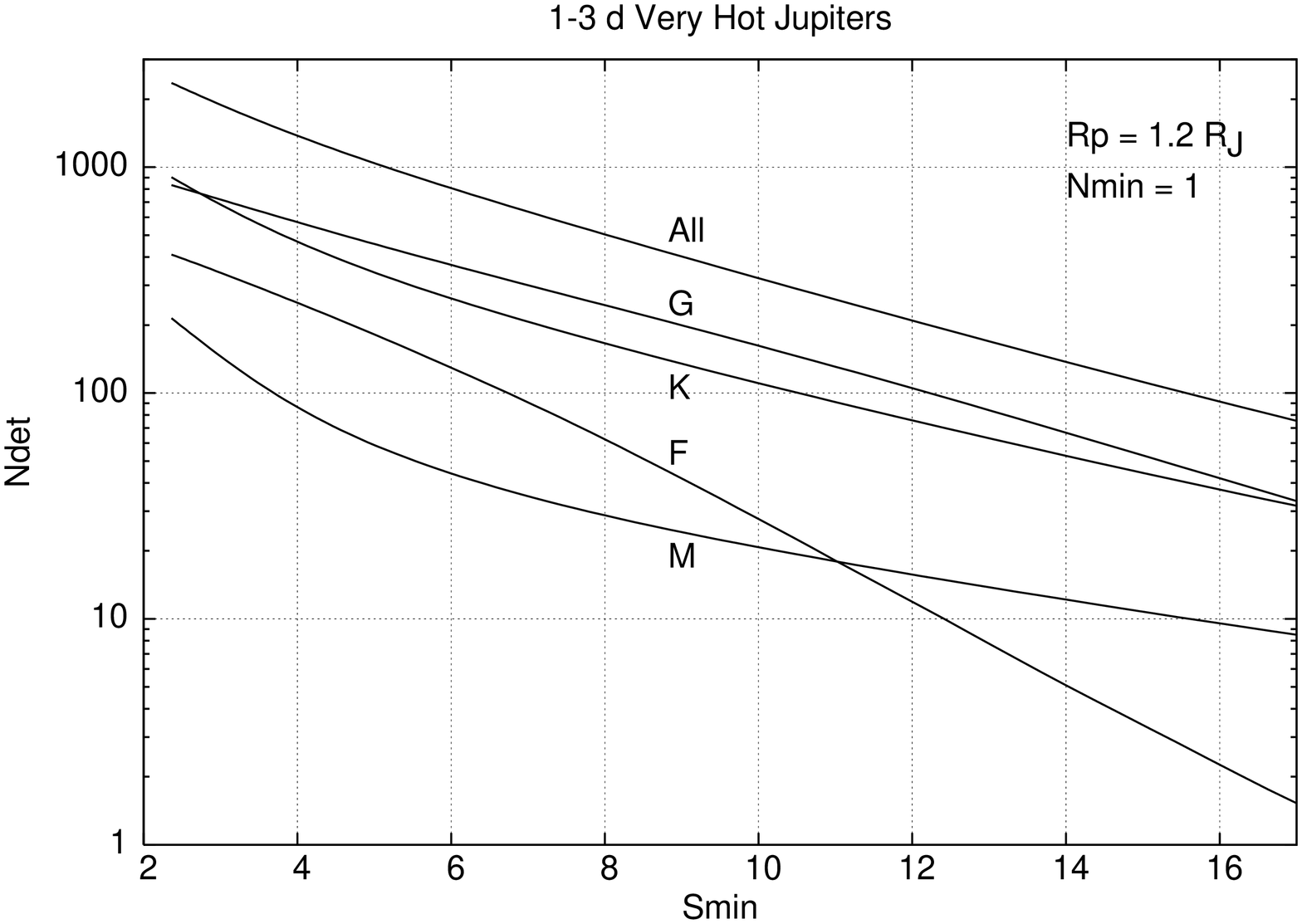,angle=270.0,width=0.5\linewidth} \label{fig:Ndet_vs_thresh_P1to3}} \\
\subfigure[]{\epsfig{file=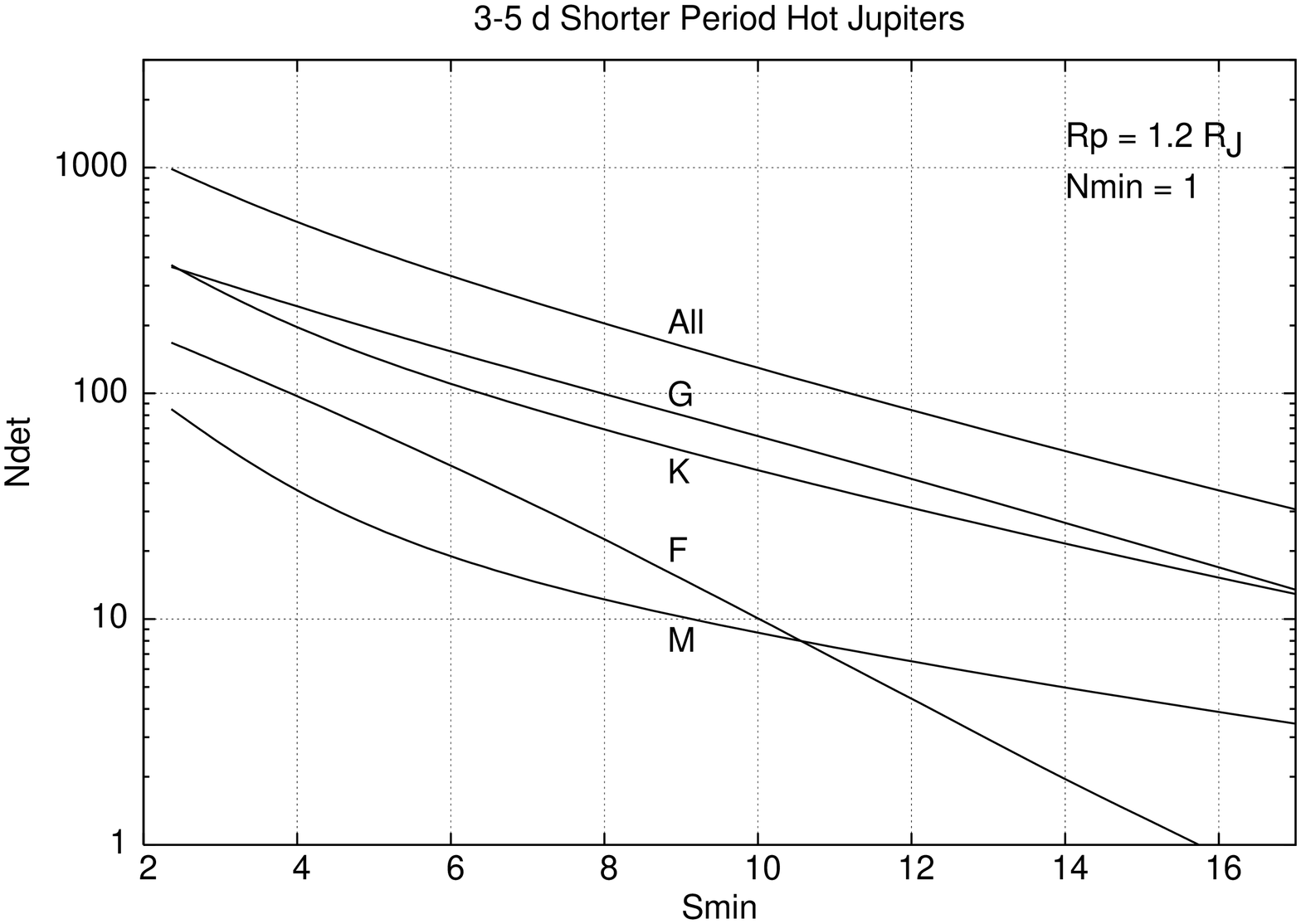,angle=270.0,width=0.5\linewidth} \label{fig:Ndet_vs_thresh_P3to5}} &
\subfigure[]{\epsfig{file=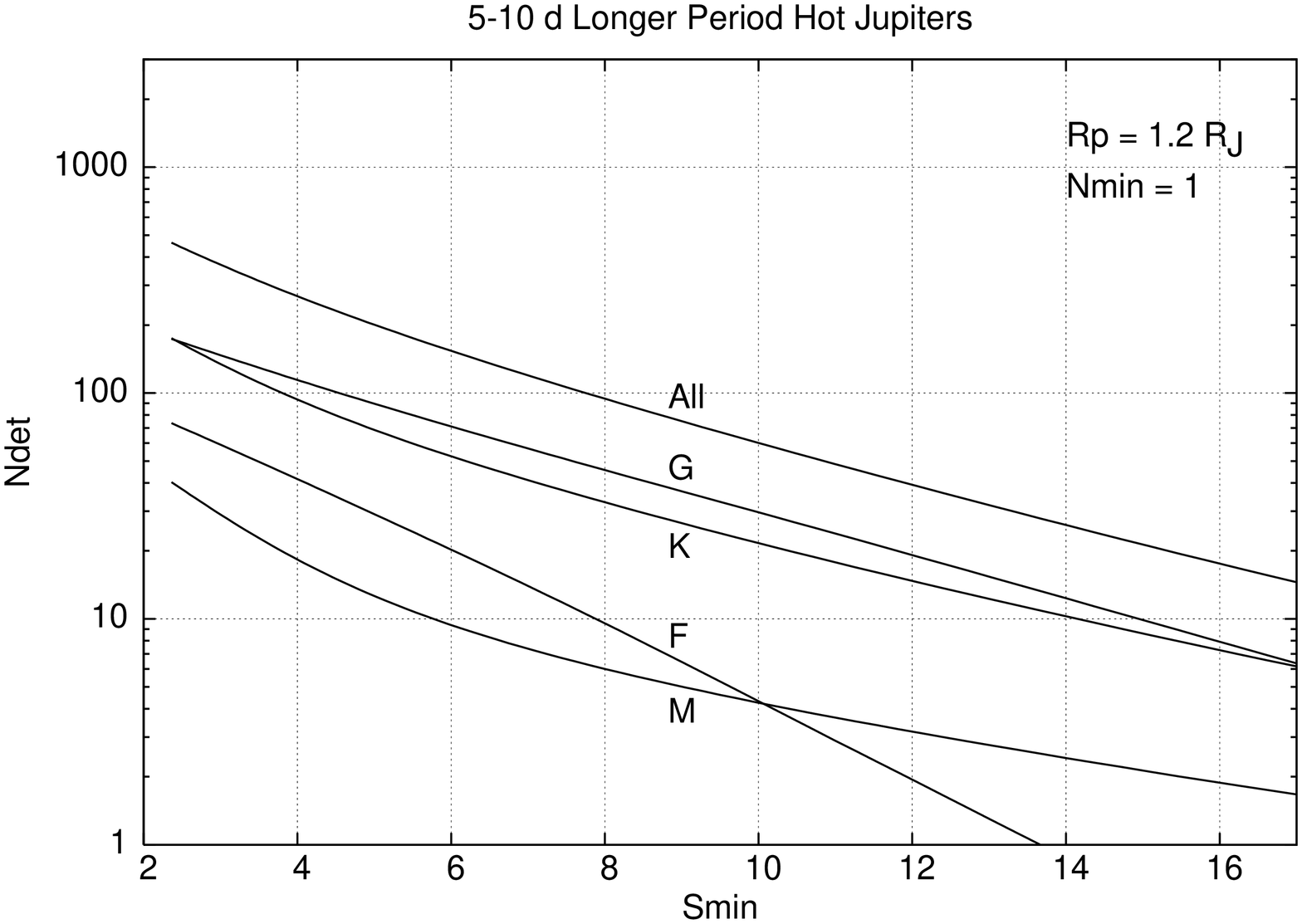,angle=270.0,width=0.5\linewidth} \label{fig:Ndet_vs_thresh_P5to10}} \\
\end{tabular}
\caption{(a): Expected number of transiting hot Jupiter detections (continuous curves) and false alarms (dashed curves) for all stars
         as functions of $S_{\mbox{\scriptsize min}}$ with $N_{\mbox{\scriptsize min}}~=~1$ and
         $R_{\mbox{\scriptsize p}} = 1.2 R_{\mbox{\scriptsize J}}$. Each curve is labelled with the hot Jupiter
         period range to which it corresponds.
         (b): Expected number of transiting 1-3~d hot Jupiter detections as a function of
         $S_{\mbox{\scriptsize min}}$ with $N_{\mbox{\scriptsize min}}~=~1$ and
         $R_{\mbox{\scriptsize p}} = 1.2 R_{\mbox{\scriptsize J}}$ for various sets of stars. Each curve is 
         labelled with the star type to which it corresponds.
         (c): Expected number of transiting 3-5~d hot Jupiter detections as a function of
         $S_{\mbox{\scriptsize min}}$ with $N_{\mbox{\scriptsize min}}~=~1$ and
         $R_{\mbox{\scriptsize p}} = 1.2 R_{\mbox{\scriptsize J}}$ for various sets of stars. 
         Each curve is labelled with the star type to which it corresponds.
         (d): Expected number of transiting 5-10~d hot Jupiter detections as a function of
         $S_{\mbox{\scriptsize min}}$ with $N_{\mbox{\scriptsize min}}~=~1$ and
         $R_{\mbox{\scriptsize p}} = 1.2 R_{\mbox{\scriptsize J}}$ for various sets of stars. 
         Each curve is labelled with the star type to which it corresponds.
\label{fig:hotjup}}
\end{figure*}

Assuming that each star S has one planet of radius $R_{\mbox{\small p}}$ and period $P$, then
the expected number of transiting planet detections
$N_{\mbox{\small det}}
\left(Y,S_{\mbox{\small min}},N_{\mbox{\small min}},R_{\mbox{\small p}},P\right)$
as a function of star type $Y$, $S_{\mbox{\small min}}$, $N_{\mbox{\small min}}$,
$R_{\mbox{\small p}}$ and $P$ is simply the sum of the detection probabilities for all stars of the
required type:
\begin{multline}
N_{\mbox{\small det}}
\left(Y,S_{\mbox{\small min}},N_{\mbox{\small min}},R_{\mbox{\small p}},P\right) \\ =
\sum_{\text{S} \in Y}
\text{P}
\left(\text{det}\,|\,\text{S},S_{\mbox{\small min}},N_{\mbox{\small min}},R_{\mbox{\small p}},P\right)
\label{eqn:Ndet}
\end{multline}  
Similarly, the expected number of false alarms
$N_{\mbox{\small fal}}
\left(Y,S_{\mbox{\small min}},N_{\mbox{\small min}},R_{\mbox{\small p}},P\right)$
is given by:
\begin{multline}
N_{\mbox{\small fal}}
\left(Y,S_{\mbox{\small min}},N_{\mbox{\small min}},R_{\mbox{\small p}},P\right) \\ =
\sum_{\text{S} \in Y}
\text{P}
\left(\text{fal}\,|\,\text{S},S_{\mbox{\small min}},N_{\mbox{\small min}},R_{\mbox{\small p}},P\right)
\label{eqn:Nfal}
\end{multline}  

In Fig.~\ref{fig:NdetNfal_vs_P}, we plot the expected number of transiting planet detections (Fig.~\ref{fig:Ndet_vs_P}) and
the expected number of false alarms (Fig. \ref{fig:Nfal_vs_P}) for all stars as functions of the period with
$S_{\mbox{\small min}}~=~10$ and $R_{\mbox{\small p}} = 1.2 R_{\mbox{\small J}}$. These quantities have
a strong dependence on period in the same way as the probabilities from which they are derived (see
Section 4.2). Fig. \ref{fig:Nfal_vs_P} clearly shows that, by increasing $N_{\mbox{\small min}}$
from 1 to 2, the expected number of false alarms is effectively reduced to zero for all $P$. However,
introducing this extra constraint for a transit detection reduces the expected yield of planets
from the survey by more than a factor of 2 (Fig. \ref{fig:Ndet_vs_P}). Also, setting $N_{\mbox{\small min}}~=~2$ is unnecessary since
the expected number of false alarms for the detection threshold
chosen in BRA05 ($S_{\mbox{\small min}}~=~10$ and $N_{\mbox{\small min}} = 1$) is less than 1 for
$P > 1.34$~d, indicating a good choice of detection threshold for all except the shortest period hot Jupiters.

\subsection{Expected Number Of Transiting Hot Jupiter Detections}

We can now estimate the expected number of detections and false alarms for three different planet period ranges,
and for different star types. We consider the very hot Jupiters with periods of
1-3~d, the shorter period hot Jupiters with periods of 3-5~d and the longer period hot Jupiters with periods
of 5-10~d. Within each period range 
$P_{1}~\leq~P~\leq~P_{M}$ we assume that planets are uniformly distributed in $\log(P)$. This is roughly 
consistent with the results of the radial velocity surveys (\citealt{hea99}; \citealt{thesis}).

Assuming that each star S has one planet of radius $R_{\mbox{\small p}}$ in the specified period range,
then the expected number of transiting planet detections is obtained by summing over
star type $Y$ and integrating over period $P$:
\begin{multline}
N_{\mbox{\small det}}
\left(Y,S_{\mbox{\small min}},N_{\mbox{\small min}},R_{\mbox{\small p}},P_{1},P_{M}\right) \\ =
\sum_{\text{S} \in Y}
\int_{P_{1}}^{P_{M}}
\left( \frac{\text{d}\ln P}{\ln \left( P_{M} / P_{1} \right)} \right)
\text{P}
\left(\text{det}\,|\,\text{S},S_{\mbox{\small min}},N_{\mbox{\small min}},R_{\mbox{\small p}},P\right)
\label{eqn:Ndet_fin}
\end{multline}
Similarly, the expected number of false alarms is given by:
\begin{multline}
N_{\mbox{\small fal}}
\left(Y,S_{\mbox{\small min}},N_{\mbox{\small min}},R_{\mbox{\small p}},P_{1},P_{M}\right) \\ =
\sum_{\text{S} \in Y}
\int_{P_{1}}^{P_{M}}
\left( \frac{\text{d}\ln P}{\ln \left( P_{M} / P_{1} \right)} \right)
\text{P}
\left(\text{fal}\,|\,\text{S},S_{\mbox{\small min}},N_{\mbox{\small min}},R_{\mbox{\small p}},P\right)
\label{eqn:Pfal_fin}
\end{multline}

\begin{table*}
\centering
\caption{Results of the Monte Carlo simulations for $S_{\mbox{\scriptsize min}} = 10$,
         $N_{\mbox{\scriptsize min}} = 1$ and
         $R_{\mbox{\scriptsize p}} = 1.2 R_{\mbox{\scriptsize J}}$.
         The number in brackets for $N_{\mbox{\scriptsize det}}$ and $N_{\mbox{\scriptsize fal}}$ 
         indicates the uncertainty on the last decimal place. 
         The upper limits on $f_{\mbox{\scriptsize p}}$ at the significance level $\alpha$ may be scaled
         with $S_{\mbox{\scriptsize min}}$ and $R_{\mbox{\scriptsize p}}$ 
         by using the scaling relation defined in Eqn. \ref{eqn:empirical5} together with the listed values of $A$.
         \label{tab:limits}}
\begin{tabular}{ccccccccc}
\hline
$Y$ & No. Of & Period &
$N_{\mbox{\scriptsize det}}$ &
$N_{\mbox{\scriptsize fal}}$ & $A$ &
Upper Limit On $f_{\mbox{\scriptsize p}}$ &
Upper Limit On $f_{\mbox{\scriptsize p}}$ &
Upper Limit On $f_{\mbox{\scriptsize p}}$ \\
 & Stars & Range & & & &
At $\alpha = 0.50$ &
At $\alpha = 0.05$ &
At $\alpha = 0.01$ \\
\hline
All Stars & 32027 & $1\text{d} \leq P \leq 3\text{d}$ & 325.09(2) & 0.792(1) & 2.18 & 0.213\% & 0.922\% & 1.42\% \\
All Stars & 32027 & $3\text{d} \leq P \leq 5\text{d}$ & 130.77(2) & 0.267(1) & 2.17 & 0.530\% & 2.29\% & 3.52\% \\
All Stars & 32027 & $5\text{d} \leq P \leq 10\text{d}$ & 60.59(1) & 0.115(1) & 2.15 & 1.14\% & 4.94\% & 7.60\% \\
\hline
Late F Stars & 3129 & $1\text{d} \leq P \leq 3\text{d}$ & 28.50(1) & 0.223(1) & 4.20 & 2.43\% & 10.5\% & 16.2\% \\
Late F Stars & 3129 & $3\text{d} \leq P \leq 5\text{d}$ & 10.37(1) & 0.075(1) & 4.10 & 6.68\% & 28.9\% & 44.4\% \\
Late F Stars & 3129 & $5\text{d} \leq P \leq 10\text{d}$ & 4.38(1) & 0.028(1) & 3.99 & 15.8\% & 68.4\% & 105\% \\
\hline
G Stars & 7423 & $1\text{d} \leq P \leq 3\text{d}$ & 163.57(2) & 0.294(1) & 2.18 & 0.424\% & 1.83\% & 2.82\% \\
G Stars & 7423 & $3\text{d} \leq P \leq 5\text{d}$ & 65.37(1) & 0.110(1) & 2.20 & 1.06\% & 4.58\% & 7.04\% \\
G Stars & 7423 & $5\text{d} \leq P \leq 10\text{d}$ & 29.85(1) & 0.051(1) & 2.19 & 2.32\% & 10.0\% & 15.4\% \\
\hline
K Stars & 15381 & $1\text{d} \leq P \leq 3\text{d}$ & 111.76(1) & 0.265(1) & 1.91 & 0.620\% & 2.68\% & 4.12\% \\
K Stars & 15381 & $3\text{d} \leq P \leq 5\text{d}$ & 46.15(1) & 0.078(1) & 1.94 & 1.50\% & 6.49\% & 9.98\% \\
K Stars & 15381 & $5\text{d} \leq P \leq 10\text{d}$ & 21.80(1) & 0.035(1) & 1.94 & 3.18\% & 13.7\% & 21.1\% \\
\hline
M Stars & 6094 & $1\text{d} \leq P \leq 3\text{d}$ & 21.26(1) & 0.010(1) & 1.44 & 3.26\% & 14.1\% & 21.7\% \\
M Stars & 6094 & $3\text{d} \leq P \leq 5\text{d}$ & 8.88(1) & 0.004(1) & 1.51 & 7.81\% & 33.7\% & 51.9\% \\
M Stars & 6094 & $5\text{d} \leq P \leq 10\text{d}$ & 4.28(1) & 0.002(1) & 1.51 & 16.2\% & 70.0\% & 108\% \\
\hline
\end{tabular}
\end{table*}

In Fig. \ref{fig:Ndet_vs_thresh_Allstar} we plot the expected number of transiting planet detections (continuous curves) and false alarms 
(dashed curves) for all stars as functions of $S_{\mbox{\small min}}$ with $N_{\mbox{\small min}}~=~1$ 
and $R_{\mbox{\small p}} = 1.2 R_{\mbox{\small J}}$. Each curve is labelled with the 
period range to which it corresponds. Similarly, each of Figs. \ref{fig:Ndet_vs_thresh_P1to3}~-~(d) corresponds to a different period range in 
which we plot the expected number of transiting planet detections for the F, 
G, K and M stars in our sample as a function of $S_{\mbox{\small min}}$ with $N_{\mbox{\small min}}~=~1$
and $R_{\mbox{\small p}} = 1.2 R_{\mbox{\small J}}$. 

For a limited range $7 \le S_{\mbox{\small min}} \le 15$, the ``curves'' for $N_{\mbox{\small det}}$ in all of 
Figs. 5(a)-(d) can be approximated by straight lines, allowing us to express our results in the form of an 
approximate empirical relationship: 
\begin{equation}
\frac{ N_{\mbox{\small det}}\left(S_{\mbox{\small min}}\right) }
     { N_{\mbox{\small det}}\left(10\right) }
\approx \exp \left[ A \left( 1 - \frac{S_{\mbox{\small min}}}{10} \right) \right]
\label{eqn:empirical1}
\end{equation} 
where $A$ is a constant that may be determined for each set of stars $Y$ and period range $P_{1}$ to $P_{M}$
considered. We note that the transit detection statistic $S_{\mbox{\small tra}}$ is equivalent to the signal-to-noise
(S/N) of the fitted transit signal and that:
\begin{equation}
S_{\mbox{\small tra}} \equiv \text{S/N} \propto \frac{\Delta f}{f_{0}} \sim \left( \frac{R_{\mbox{\small p}}}{R_{*}} \right)^{2}
\label{eqn:empirical2}
\end{equation}
where $\Delta f / f_{0}$ is the fractional transit depth. 
Now, since we have $S_{\mbox{\small tra}}~\propto~R_{\mbox{\small p}}^{2}$, 
we may infer that for a fixed $S_{\mbox{\small tra}}$ the equivalent detection threshold
varies as $S_{\mbox{\small min}}~\propto~R_{\mbox{\small p}}^{-2}$. 
Using this fact, we may rewrite Eqn. \ref{eqn:empirical1} as:
\begin{equation}
\frac{ N_{\mbox{\small det}}\left( S_{\mbox{\small min}},R_{\mbox{\small p}} \right) }
     { N_{\mbox{\small det}}\left( 10, 1.2 R_{\mbox{\small J}} \right) }
\approx \exp \left[ A \left( 1 - \left( \frac{S_{\mbox{\small min}}}{10} \right)
                                 \left( \frac{1.2 R_{\mbox{\small J}}}{R_{\mbox{\small p}}} \right)^{2} \right) \right]
\label{eqn:empirical4}
\end{equation}
For $S_{\mbox{\small min}}~=~10$, the range of planetary radii over which this scaling relation is valid is
$0.98~R_{\mbox{\small J}}~\le~R_{\mbox{\small p}}~\le~1.43 R_{\mbox{\small J}}$. 

The relation in Eqn. \ref{eqn:empirical4}  
allows us to scale our results for a range of planetary radii
that encompasses the radii of the known transiting hot Jupiters without having to repeat the Monte Carlo simulations. 
In Table \ref{tab:limits} we present the results of the Monte Carlo simulations for the various 
sets of stars and orbital period
ranges considered, as calculated for $S_{\mbox{\small min}} = 10$, $N_{\mbox{\small min}} = 1$ and
$R_{\mbox{\small p}}~=~1.2~R_{\mbox{\small J}}$. We also include the value of $A$ from Eqn. \ref{eqn:empirical4}
for each combination of star and planet type.
One can see that we expect to detect $\sim$325 1-3~d very hot Jupiters,
$\sim$131 3-5~d hot Jupiters and $\sim$61 5-10~d hot Jupiters from the transit survey based on the assumption that each star 
has a single planet of the specified type.

\subsection{Upper Limits On The Hot Jupiter Fraction}

\begin{figure*}
\def\subfigtopskip{4pt}
\def\subfigbottomskip{8pt}
\def\subfigcapskip{4pt}
\centering
\begin{tabular}{cc}
\subfigure[]{\epsfig{file=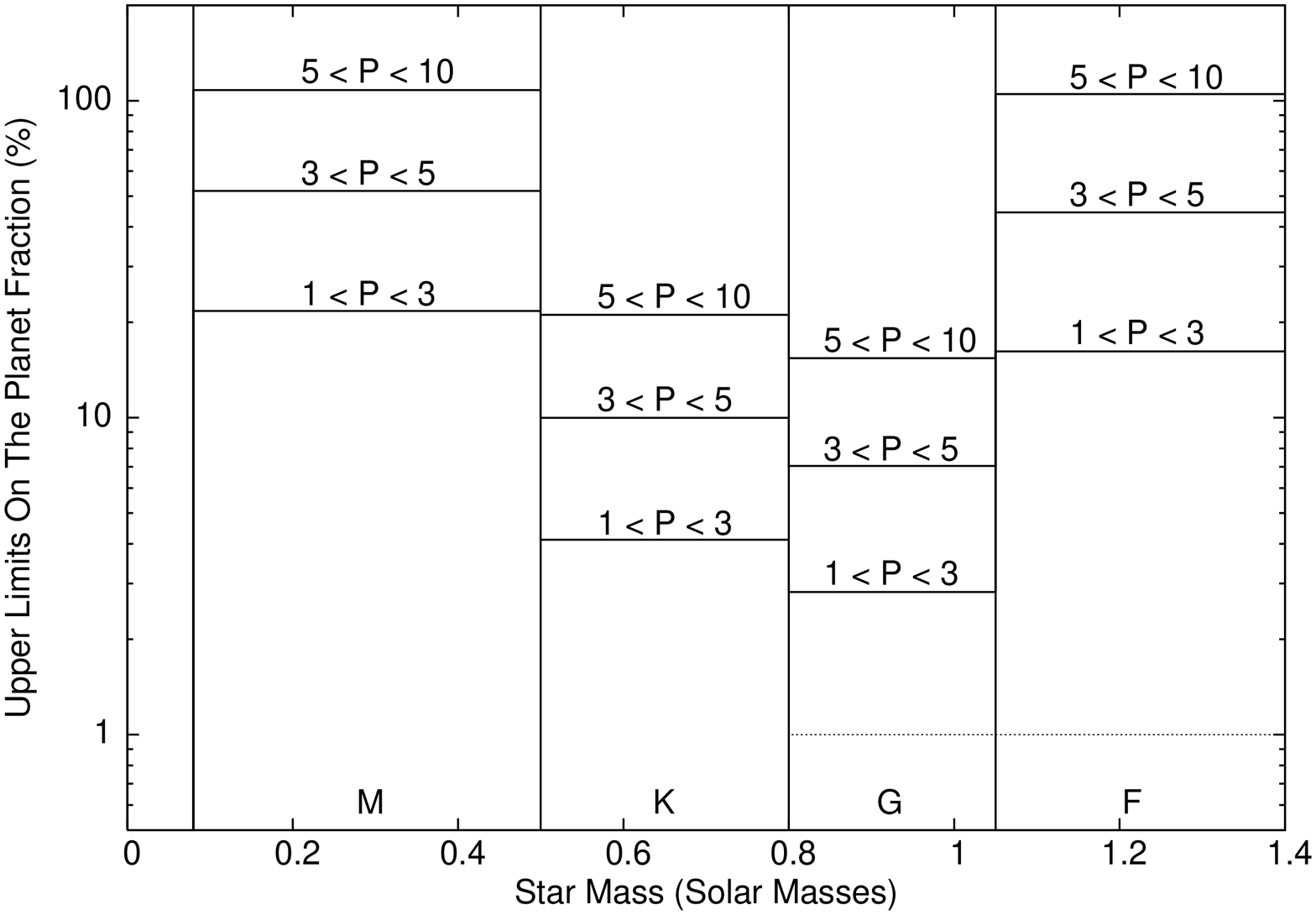,angle=270.0,width=0.5\linewidth} \label{fig:limfig1}} &
\subfigure[]{\epsfig{file=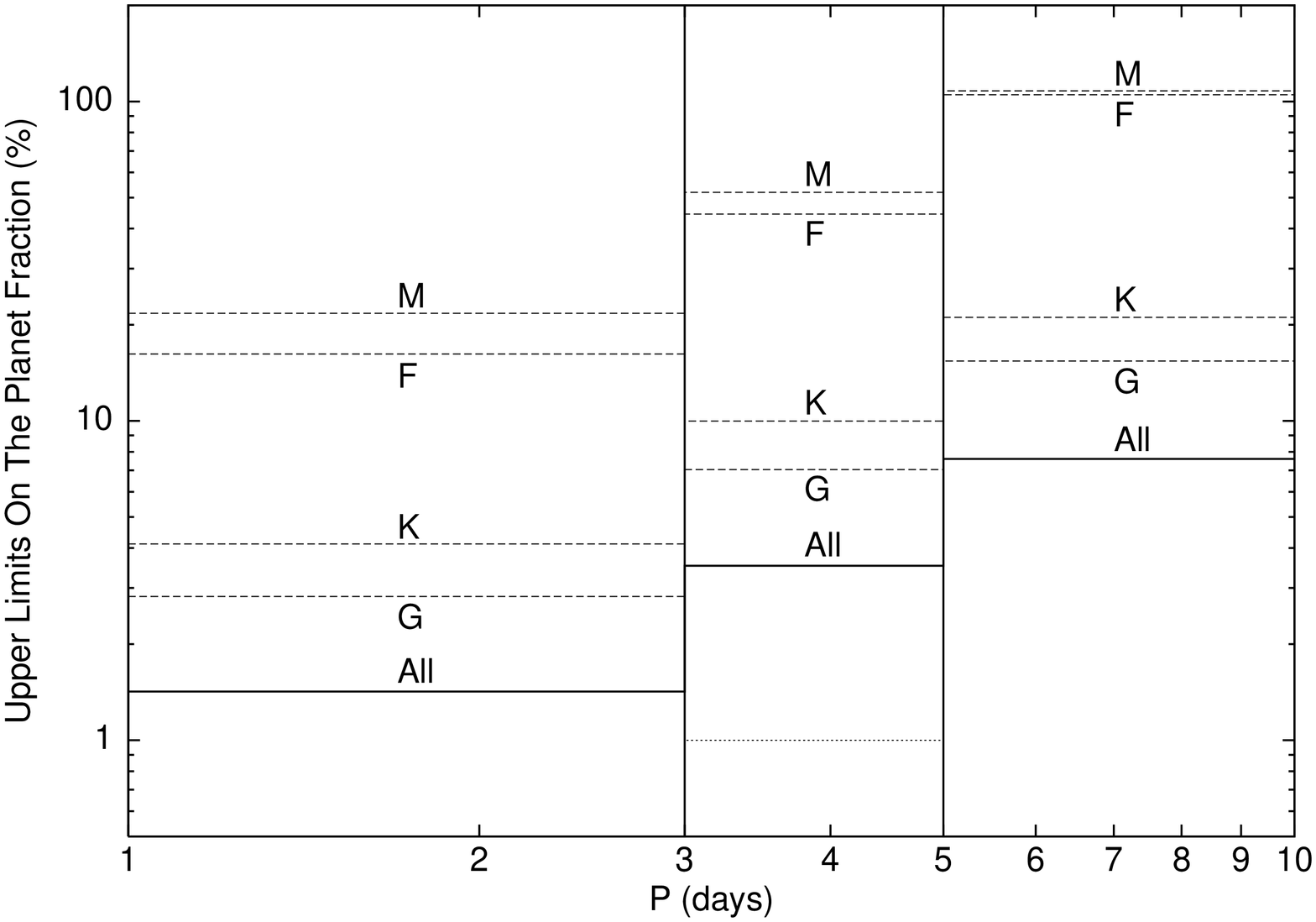,angle=270.0,width=0.5\linewidth} \label{fig:limfig2}} \\
\end{tabular}
\caption{(a): The upper limit on $f_{\mbox{\scriptsize p}}$ at $\alpha~=~0.01$ as a function of star type for the various types of hot
              Jupiters defined by their period ranges with 
              $R_{\mbox{\scriptsize p}}~=~1.2 R_{\mbox{\scriptsize J}}$. The dotted line shows the estimate of the 3-5~d
              hot Jupiter fraction for Solar neighbourhood Sun-like stars from \citet{but00}.
         (b): The upper limit on $f_{\mbox{\scriptsize p}}$ at $\alpha~=~0.01$ as a function of planet type (defined by the period range)
              with $R_{\mbox{\scriptsize p}}~=~1.2 R_{\mbox{\scriptsize J}}$ for various star types. The dotted line shows the estimate of the 3-5~d
              hot Jupiter fraction for Solar neighbourhood Sun-like stars from \citet{but00}.
\label{fig:limfig}}
\end{figure*}

In reality, only a fraction $f_{\mbox{\small p}}$ of the
stars considered in our Monte Carlo simulations harbour a planet of a specified type, instead of our
assumed one planet per star. Hence we must correct our calculations of the expected number of
transiting planet detections by this factor. We refer to $f_{\mbox{\small p}}$ as the planet fraction. Since
$f_{\mbox{\small p}}$ is an unknown quantity that we would like to estimate, we
may use the fact that the transit survey in BRA05 has most likely produced a null result
(although this is still to be confirmed) and place a significant upper limit on $f_{\mbox{\small p}}$.

First of all we make the assumption that the actual number of transiting planet detections $X$ has a Poisson
distribution with expected value $E(X)$ given by:
\begin{equation}
E(X) = f_{\mbox{\small p}} N_{\mbox{\small det}} 
\label{eqn:EX}
\end{equation}
The Poission distribution is defined by:
\begin{equation}
\text{P}(X=x) = \frac{(E(X))^{x}}{x!} e^{-E(X)} \qquad \text{for} \qquad x \in \mathbb{N}_{0} 
\label{eqn:poisson}
\end{equation}
For a null result, $x = 0$. Using this fact and substituting Eqn. \ref{eqn:EX} into Eqn. \ref{eqn:poisson} we get:
\begin{equation}
\text{P}(X=0) = e^{-f_{\mbox{\scriptsize p}} N_{\mbox{\scriptsize det}}} 
\label{eqn:PX0}
\end{equation}
In order to obtain an upper limit on $f_{\mbox{\small p}}$ at the significance level $\alpha$ we require:
\begin{equation}
\text{P}(X=0) \leq \alpha \label{eqn:condition}
\end{equation}
Hence, from Eqns. \ref{eqn:PX0} and \ref{eqn:condition}, we derive:
\begin{equation}
f_{\mbox{\small p}} \leq \frac{- \ln \alpha}{N_{\mbox{\small det}}} 
\label{eqn:upperlim} 
\end{equation}
The values of $f_{\mbox{\small p}}$ that we derive in this manner for $\alpha = 0.50$, $\alpha = 0.05$ and $\alpha = 0.01$ are shown in
Table \ref{tab:limits}. For example, for $\alpha~=~0.01$, we place an upper limit of 1.42\% on the 1-3~d very hot Jupiter fraction
based on the assumption that such planets have a typical radius of $1.2~R_{\mbox{\small J}}$. Our most conservative
upper limits are obtained for $\alpha = 0.01$ and consequently these are the upper limits on $f_{\mbox{\small p}}$ that we
consider in the discussion and conclusions.

Finally, we derive how the upper limit on $f_{\mbox{\small p}}$ scales with $S_{\mbox{\small min}}$ and $R_{\mbox{\small p}}$ by 
using Eqn. \ref{eqn:empirical4} in Eqn. \ref{eqn:upperlim}:
\begin{equation}
f_{\mbox{\small p}} \leq 
\frac{- \ln \alpha}{N_{\mbox{\small det}}\left( 10, 1.2 R_{\mbox{\small J}} \right)}
\exp \left[ A \left( \left( \frac{S_{\mbox{\small min}}}{10} \right)
                     \left( \frac{1.2 R_{\mbox{\small J}}}{R_{\mbox{\small p}}} \right)^{2} - 1 \right) \right]
\label{eqn:empirical5}
\end{equation}

\section{Discussion}

We have derived relatively stringent upper limits on the abundance of hot Jupiters for the field of NGC~7789.
The most stringent upper limit on $f_{\mbox{\small p}}$ at $\alpha~=~0.01$ of 0.79\% is obtained for 1-3~d very hot Jupiters of radius 
$1.4~R_{\mbox{\small J}}$.
For the Sun-like G stars, we obtain limits on $f_{\mbox{\small p}}$ at $\alpha~=~0.01$ of 2.82\% for the 1-3~d very hot Jupiters and 7.04\% for the
3-5~d hot Jupiters (assuming $R_{\mbox{\small p}} = 1.2 R_{\mbox{\small J}}$).
Fig.~\ref{fig:limfig} shows plots of the upper limit on the planet fraction $f_{\mbox{\small p}}$ at a significance level
of 1\% as a function of star type (Fig.~\ref{fig:limfig1}) and orbital period (Fig.~\ref{fig:limfig2}) for an assmued planetary radius of
$1.2~R_{\mbox{\small J}}$. We also show the estimate by \citet{but00}
that $\sim$1\% of nearby Sun-like stars (late F and G dwarfs) host a 3-5~d hot Jupiter, as derived from radial velocity observations
(dotted line).

It is interesting to note that although the K stars are the most numerous in our star sample (15381 stars), it is the
7423 G stars that produce the largest expected number of transiting planets, and therefore the strictest upper limits on
$f_{\mbox{\small p}}$ (Table \ref{tab:limits}). This is due to the fact that the G stars are generally brighter than the K stars in our sample,
and therefore the corresponding gain in accuracy of the photometric measurements outweighs the smaller number of stars for which we
can search for transits and the smaller transit signal for a given planetary radius.

We may compare our results directly with those of \citet{but00} by considering the late F and G stars in our sample and
the corresponding expected number of transiting planets for the 3-5~d hot Jupiters with 
$R_{\mbox{\small p}} = 1.2 R_{\mbox{\small J}}$. We expected to detect $\sim$10.4 such
planets around the late F stars in our sample, and $\sim$65.4 such planets around the G stars in our sample, an expected total of
$\sim$75.7 3-5~d hot Jupiters. This places an upper limit on $f_{\mbox{\small p}}$ of $\sim$6.08\%
at the 1\% significance level for these types of
star and planet. This is consistent with the estimate of the hot Jupiter fraction derived by \citet{but00} of 
$f_{\mbox{\small p}} \approx$~1\% and
demonstrates with confidence that the hot Jupiter fraction for Sun-like stars in this field may not be more than a factor
of $\sim$6 times greater than that for the Solar neighbourhood. 

By considering the number of planets detected by the radial-velocity technique, and by the transit technique from the OGLE survey,
\citet{gau05} estimate that the relative frequency of 1-3~d very hot Jupiters to 3-9~d hot Jupiters is $0.18^{+0.12}_{-0.08}$. Using this 
result together with $f_{\mbox{\small p}}~\approx$~1\% for hot Jupiters, the authors calculate that 
$f_{\mbox{\small p}}~\approx~0.1\%-0.2\%$ for 1-3~d very hot Jupiters.
This is consistent with our upper limit on $f_{\mbox{\small p}}$ at $\alpha~=~0.01$ of 1.42\% for 1-3~d very hot Jupiters of radius 
$1.2~R_{\mbox{\small J}}$.

\section{Conclusions}

The calculation of the expected number of transiting planet detections for a transit survey has been discussed
in varying levels of detail by several authors (e.g: \citealt{gil00}; \citealt{wel05}; \citealt{hid05}; \citealt{hoo05}; \citealt{moc05}). 
A requisite for such a calculation is an estimate of the masses and radii of the stars in the sample from observational constraints or a model for the 
star population that predicts these properties. 

In this paper, we present a method for calculating in detail the detection probabilities
(and false alarm rates) of transiting planets for photometric time-series data as functions of detection threshold,
planetary radius and orbital period. The calculation is based on Monte Carlo
simulations and requires the properties of the host stars to be known, either from observational constraints as in the case of stellar
clusters or a model for the star population. We have shown how to convert these probabilities into an expected number of transiting
planet detections as a function of star and planet type. For a null result in a transit survey, this information can be used to determine
a significant upper limit on the planet fraction.

In the case of the transit survey of NGC~7789 presented in BRA05 we have derived upper limits on the planet fraction for the F, G, K and M
stars in the sample and for three relevant period ranges of hot Jupiters. We have also derived how these limits scale with detection threshold
and planetary radius.
In BRA05, it is estimated that the survey expects to detect $\sim$2 HD 209458b-like transiting planets or $\sim$4 OGLE-TR-56b-like transiting 
planets using simple arguments. Our results indicate that for HD 209458b (3-5~d hot Jupiter with 
$R_{\mbox{\small p}}~\sim~1.4~R_{\mbox{\small J}}$), and under the assumption that $f_{\mbox{\small p}} \approx$~1\%, we also expect to detect 
$\sim$2 such transiting planets. Similarly for OGLE-TR-56b (1-3~d very hot Jupiter with
$R_{\mbox{\small p}}~\sim~1.2~R_{\mbox{\small J}}$), and again under the assumption that $f_{\mbox{\small p}} \approx$~1\%, we expect to detect
$\sim$3 such transiting planets. It is encouraging to note the agreement between the two methods although the simpler method from BRA05 may tend to 
slightly over estimate the planetary detection rate. We conclude that the transit survey presented in BRA05 reached the sensitivity 
required in order to detect a few hot Jupiters if the abundance of such planets
in the field of NGC~7789 is similar to that of the Solar neighbourhood.

Improved survey design, mainly by employing a longer survey duration, will greatly improve the sensitivity to hot Jupiters.
It is well known that metal rich stars have a much higher probability of hosting an extra-solar planet (\citealt{san04}) and hence careful 
choice of 
the target star population will increase the probability of a detection. Even in the presence of a null result, a more sensitive survey will 
allow the derivation of tighter limits on the planet abundance.

\section*{Acknowledgements}

The calculations presented in this work were done using IDL programs and the CONDOR distributed computing software.
IDL is provided, under license, by Research Systems Inc. DMB is grateful to PPARC for the provision of a PhD studentship
whilst at St. Andrews University and post-doctoral support at LJMU, the latter as part
of the ``RoboNet-1.0'' project.

\label{lastpage}

\end{document}